# Mining Top-K Co-Occurrence Items

Zhi-Hong Deng

*Abstract*—Frequent itemset mining has emerged as a fundamental problem in data mining and plays an important role in many data mining tasks, such as association analysis, classification, etc. In the framework of frequent itemset mining, the results are itemsets that are frequent in the whole database. However, in some applications, such recommendation systems and social networks, people are more interested in finding out the items that occur with some user-specified itemsets (*query itemsets*) most frequently in a database. In this paper, we address the problem by proposing a new mining task named *top-k co-occurrence item mining*, where *k* is the desired number of items to be found. Four baseline algorithms are presented first. Then, we introduce a special data structure named *Pi-Tree* (*Prefix itemset Tree*) to maintain the information of itemsets. Based on *Pi-Tree*, we propose two algorithms, namely *PT* (*Pi-Tree-based algorithm*) and *PT-TA* (*Pi-Tree-based algorithm with TA pruning*), for mining top-*k* co-occurrence items by incorporating several novel strategies for pruning the search space to achieve high efficiency. The performance of *PT* and *PT-TA* was evaluated against the four proposed baseline algorithms on both synthetic and real databases. Extensive experiments show that *PT* not only outperforms other algorithms substantially in terms execution time but also has excellent scalability.

## I. INTRODUCTION

Frequent itemset mining was first proposed by Agrawal et al. [2] for market basket analysis in dealing with the problem of mining association rules. Since the first proposal of this data mining task and its associated efficient mining algorithms, there have been thousands of follow-up research publications on various kinds of extensions and applications. Frequent itemset mining has emerged as an important topic in data mining. It has been proven to play an essential role in many data mining tasks such as mining associations, correlations, causality, sequential itemsets, episodes, multi-dimensional itemsets, max-itemsets, partial periodicity and emerging itemsets [17].

With frequent itemset mining, an itemset is frequent if its occurrence frequency in a database is not less than a given threshold. That is, frequent itemset is a global concept in terms of the whole database without relevant to any special itemsets.

This work was supported in part by the National Natural Science Foundation of China (Grant No. 61170091) and the National High Technology Research and Development Program of China (Grant No. 2009AA01Z136).

Z. H. Deng is with Key Laboratory of Machine Perception (Ministry of Education), Department of Machine Intelligence, School of Electronics Engineering and Computer Science, Peking University, 100871 Beijing, China (e-mail: zhdeng@cis.pku.edu.cn).

However, in some application, such as recommendation systems [12, 29] and social networks [21], people may more concern these items most relevant to some query itemsets. For example, in a recommendation system of an online shop on the WWW, whenever a user purchases some products, the system should instantly recommend other products that have been bought together with these products to sell more products. Such task can be formally illustrated by the following example.

**Example 1** Let us consider a simplified database *DB* on shopping transactions as shown in Table 1, where TID and Items are attributes with domains {1, 2, 3, 4, 5} and {*a*, *b*, *c*, *d*, *e*, *f*, *g*} respectively. Each transaction *T* in database *DB* has a unique identifier (TID) and is a subset of {*a*, *b*, *c*, *d*, *e*, *f*, *g*}. For simplicity, we represent the transaction with *tid* = *j* by $T_j$. For example, $T_2$ stands for the second transaction in Table 1. The top-2 items co-occurring with itemset {*ac*} are item *f* and *d* as three shopping transactions contain {*acf*} and two shopping transactions contain {*acd*} while other items occur with {*ac*} at most in one shopping transaction. Similarly, the top-2 items co-occurring with itemset {*c*} are item *a* and *f*.

**Table 1. A database as running example**

| TID | Items |
|---|---|
| 1 | *b*, *f*, *g* |
| 2 | *a*, *b*, *c*, *f* |
| 3 | *a*, *c*, *d*, *f* |
| 4 | *b*, *c*, *e* |
| 5 | *a*, *c*, *d*, *e*, *f* |

Although many algorithms have been proposed for mining frequent itemsets [16] and top-*k* frequent itemsets [18], it is not possible to simply adapt these algorithms to mining top-*k* co-occurrence items. The reason can be stated as follows. First, mining top-*k* co-occurrence items has no the requirement of setting minimal support threshold as mining frequent itemsets or top-*k* frequent itemsets, where a frequent itemsets is defined as the one whose support is no less than the minimal support threshold. Therefore, the algorithms for mining frequent itemsets or top-*k* frequent itemsets can not be directly used to mining top-*k* co-occurrence items since it is infeasible to set an appropriate threshold for finding the top-*k* co-occurrence items of each itemset. Note that, the minimal supports of the top-*k* co-occurrence items of different itemsets are varied and we can not know them in advance. Second, the top-*k* co-occurrence items do not have the downward closure property [2], which is the foundation of all algorithms for mining frequent itemsets or top-*k* frequent itemsets.

For a given query itemset, a naive approach for finding its

top-*k* co-occurrence items is to exhaustively examine all items that do not belong to the itemsets from the database. Obviously, this approach will encounter the large search space problem, especially when databases are very large and a large number of query itemsets need to find their top-*k* co-occurrence items. Hence, how to effectively prune the search space and efficiently find top-*k* co-occurrence items is a new challenge.

To address this issue, we systemically study the task of mining top-*k* co-occurrence items in this paper. This work has three major contributions as following.

First, we first introduce and formalize the problem of top-k co-occurrence item mining. As far as we know, the problem has not been previously proposed and studied.

Second, six algorithms are proposed for mining top-*k* co-occurrence items. The first four algorithms, which are *NT* (*Naive Hunting algorithm*), *NTI* (*Naive Hunting algorithm with Inverted list index*), *NT-TA* (*NT with TA pruning*), and *NTI-TA* (*NTI with TA pruning*) respectively, are baseline methods. NT discovers top-*k* co-occurrence items by directly scanning the original database. NTI finds top-*k* co-occurrence items by employing Tid-sets [37], a kind of inverted list, to prune the search space. NT-TA and NTI-TA are corresponding extensions of NT and NTI by using TA strategy to prune the search space. The last two algorithms, namely PT and PT-TA, are based on Pi-Tree. The Pi-Tree structure provides a good platform for designing novel strategies for pruning the search space greatly to achieve high efficiency. PT-TA is the extension of PT by using *TA* strategy to prune the search space.

Third, we conducted substantial experiments on both synthetic and real databases to evaluate the performance of the six algorithms. The experimental results show that PT outperforms other algorithms in terms of execution time, especially when the databases are dense. In addition, PT also shows excellent scalability.

The rest of this paper is organized as follows. In Section 2, we survey the related work. In Section 3, we introduce the problem of top-*k* co-occurrence itemset mining. In Section 4, the proposed data structures and algorithms are described in details. Experimental results and evaluation are presented in Section 5. Finally, our work is summarized in Section 6.

## II. RELATED WORK

The problem of frequent itemset mining is to find the complete set of itemsets satisfying a given minimum support in a transaction database. Many studies have been proposed for mining frequent itemsets [1, 13, 16, 41]. All these studies use a property named *downward closure* to prune the infrequent itemsets. This property indicates that if an itemset is infrequent, then all of its supersets must be infrequent. One of the well-known algorithms is Apriori algorithm [3, 23, 27]. Apriori algorithm employ candidate set generation-and-test strategy to discover frequent itemsets. Although Apriori algorithm achieves good performance, some studies reveal that it needs several whole database scans and suffers from a large number of candidates. In order to avoid scanning whole database repeatedly, a number of vertical mining algorithms have been proposed [4, 28, 37, 38, 39]. Zaki employed Tid-sets [37] and diffsets [38] to represent itemsets. The counting of supports of itemsets can be obtained via intersection of itemsets' Tid-sets or difference of itemsets' diffsets, which avoids scanning the whole database. Thereafter, Deng et al. proposed a novel framework for representation of itemset, where each itemset is represented by a set of nodes in a prefix tree. Based on such representation framework, some algorithms [8, 9, 10, 11, 32, 33], such Prepost, were proposed to mine frequent itemsets. Prepost [10] adopts N-Lists to represent itemsets. It is more efficient than those algorithms based on Tid-sets or diffset, such as Eclat [37] and dEclat [38] in that N-Lists are much more compressed than Tid-sets or diffsets [10]. In order to solve the problem of huge number of candidates generated by generation-and-test strategy, FP-growth-based algorithms [15, 17, 24, 26, 30] adopts a highly condensed data structure called FP-tree to store databases and employs a partitioning-based, divide-and-conquer approach to find frequent itemsets without any candidate generation. In recent years, some parallel algorithms [25, 36] have been proposed for mining frequent itemsets since datasets in modern data mining applications become excessively large.

One problem, which hampers the popular use of frequent itemset mining, is that it is difficult for users to choose a proper minimum support threshold. To accurately control the number of the most frequent itemsets, a good solution is to mine top-*k* frequent itemsets [5, 6, 7, 18, 20, 22, 31, 34, 40] instead of mine frequent itemsets. Setting *k* is more intuitive than setting the minimal support threshold because *k* represents the number of itemsets that the user wants to find whereas choosing the threshold depends solely on database's characteristics, which are often unknown to users [35]. Although these algorithms are distinct in that they employ different data structures and search strategies to discover top-*k* frequent itemsets, they follow a same general process for discovering top-*k* itemsets. The main idea of the general process is to maintain a list of itemsets *L* and a minimum support threshold *minsup*. *L* contains the *Top-k* frequent itemsets that have been found so far. By raising the value of *minsup* to the minimal value of supports of all itemsets in *L*, the search space can be greatly pruned when searching for more itemsets.

Different from frequent itemsets and top-*k* frequent itemsets, top-*k* co-occurrence items need not to satisfy a given minimal support threshold and do not have the downward closure property. Therefore, the techniques for frequent itemset mining or top-*k* frequent itemset mining can not be directly applied to mine top-*k* co-occurrence items. As far as we know, no previous work has proposed or studied the problem of mining top-*k* co-occurrence items.

## III. PROBLEM FORMULATION

In this section, we formally define the problem of top-*k* co-occurrence itemset mining. Let $I = \{i_1, i_2, \ldots, i_m\}$ be the universal item set. Let $DB = \{T_1, T_2, \ldots, T_n\}$ be a transaction database, where each $T_k$ ($1 \leq k \leq n$) is a transaction which is a set of items such that $T_k \subseteq I$. We also call $P$ an itemset if $P$ is a set of items.

**Definition 1**. Let $P$ be an itemset. A transaction $T$ is said to contain $P$ if and only if $P \subseteq T$.

**Definition 2**. The *support count* of an itemset $P$, denoted as $SC(P)$, is the number of transactions that contain $P$ in *DB*.

**Definition 3.** Given itemset $P$, an item $i$ ($i \in I \wedge i \notin P$) is a *co-occurrence item* of $P$ if and only if one or more transactions contain $P \cup \{i\}$. The set of all co-occurrence items of $P$ is denoted as $P_{coi}$.

**Definition 4.** Given itemset $P$ and $i$ ($\in P_{coi}$), the *co-occurrence count* of $P$ and $i$, denoted as $CO(P, i)$, is defined as the support count of $P \cup \{i\}$.

**Definition 5. (Top-$k$ co-occurrence item)** Given itemset $P$, an item $i$ ($\in P_{coi}$) is called a *top-k co-occurrence item* of $P$ if there are less than $k$ items whose co-occurrence counts are larger than $CO(P, i)$.

With the definitions above, we define the problem of top-$k$ co-occurrence item mining as follows.

**Problem Statement**. Given a transaction database *DB*, an itemset $P$, and the desired number $k$, the problem of finding the complete set of top-$k$ co-occurrence items of $P$ is to find $k$ items that occur most frequently with $P$ in *DB*.

Assume we want to know which items most frequently occur with $\{ac\}$ in Table 1. According to Definition 4, we know $CO(\{ac\}, b) = 1$ because only $T_2$ contains $\{abc\}$. Similarly, we have $CO(\{ac\}, d) = 2$, $CO(\{ac\}, e) = 1$, and $CO(\{ac\}, f) = 3$. If threshold $k$ is set to 1, the top-1 co-occurrence item of $\{ac\}$ is $f$. If $k$ is set to 2, the top-2 co-occurrence items of $\{ac\}$ are $f$ and $d$.

## IV. THE BASELINE ALGORITHMS

In this section, we introduce four baseline algorithms: NT, NTI, NT-TA, and NTI-TA. NT-TA and NTI-TA are corresponding extensions of NT and NTI by adopting Fagin's TA principle [14] to reach early termination after obtaining the top-k results without searching all candidates.

### A. NT and NTI Algorithms

Given an itemset $P$ and a database *DB*, NT scans every transaction in *DB*. If a transaction contains $P$, NT increases the co-occurrence count of each item $i$, not contained by $P$ in the transaction, by 1. By visiting all transactions, NT obtains co-occurrence counts of all co-occurrence items of $P$. By sorting all co-occurrence items in descending order of co-occurrence count, NT finds all top-$k$ co-occurrence items of $P$. Algorithm 1 shows the details of NT.

Obviously, NT is inefficient in that it scans the whole database where a lot of transactions do not contain $P$ and are undesirable. To avoid visiting undesirable transactions, NTI first finds all transactions containing $P$ by employing Tid-set [37], a data structure originated from inverted list. The Tid-set of an itemset (or item) is the set of TIDs of all transactions that contain the itemset (or item). As stated in [37], the Tid-set of an itemset can be computed by intersecting the Tid-sets of all items that are contained by the itemset. Then, NTI discovers the top-$k$ co-occurrence items in the projected database, which is the set of all transactions containing $P$, just as NT does. Because NTI searches the top-$k$ co-occurrence items in a smaller projected database, it is more efficient than NT. Algorithm 2 shows the details of NTI.

---

**Algorithm 1: NT Algorithm**

**Input:** Transaction database *DB*, itemset $P$, and threshold $k$.
**Output:** $R_{Tk}$, the set of all top-$k$ co-occurrence items of $P$.
1: **foreach** $T \in DB$ **do**
2:    **if** $P \subseteq T$ **then**
3:       **foreach** $i \in T - P$ **do**
4:          **if** $i$ is first visited **then**
5:             $CO(i) \leftarrow 1$;
6:          **else**
7:             $CO(i) \leftarrow CO(i) + 1$;
8: $R_{Tk} \leftarrow \{x \mid CO(x)$ is one of top $k$ biggest elements of $\{CO(i)\}\}$;
9: **return** $R_{Tk}$;

---

**Algorithm 2: NTI Algorithm**

**Input:** Transaction database *DB*, itemset $P$, and threshold $k$.
**Output:** $R_{Tk}$, the set of all top-$k$ co-occurrence items of $P$.
1: **foreach** $T \in DB$ **do**
2:    **foreach** $i \in T$ **do**
3:       **if** $i$ is first visited **then**
4:          TID_list($i$) $\leftarrow \{T.tid\}$;
5:       **else**
6:          TID_list($i$) $\leftarrow$ TID_list($i$) $\cup \{T.tid\}$;
7: TID_list($P$) $\leftarrow$ TID_list($P$.*first-item*);
8: **foreach** $i \in (P / \{P.first\text{-}item\})$ **do**
9:    TID_list($P$) $\leftarrow$ TID_list($P$) $\otimes$ TID_list($i$);
10: $DB_P \leftarrow \{T \mid T \in DB \wedge T.tid \in$ TID_list($P$)$\}$
11: run NT Algorithm by replacing *DB* with $DB_P$;

---

### B. Extensional Algorithms with TA Pruning

TA (*Threshold Algorithm*), proposed by Fagin et al [14], is extensively used in top-$k$ query processing in relational database systems. TA finds the top-$k$ objects by iteratively computing the scores of candidate objects and evaluating the lower bound score of current top-$k$ objects and the upper bound score of all remaining objects. If the lower bound score is found to be higher than the upper bound score, TA terminates the process and outputs the final top-$k$ results. In the task of mining top-$k$ co-occurrence items, we can employ the idea of TA principle to boost the mining speed. Before presenting the concrete algorithms, let's first introduce some concepts and properties.

**Definition 6.** Let $P$ be an itemset and $DB_j$ ($\subseteq DB$) be a set of transactions. The *local support count* of itemset $P$ in $DB_j$ is the number of transactions that contain $P$ in $DB_j$. For brevity, we denote it as $SC(P, DB_j)$.

**Definition 7**. Let $X$ be a set. The cardinality of $X$, denoted as $|X|$, is the number of all elements in $X$.

For example, assume $P = \{ac\}$ and $DB_j = \{T_1, T_2, T_3\}$.

According to Definition 6, $SC(\{ac\},\{T_1, T_2, T_3\})$ is 2. $|P|$ is 2 and $|DB_j|$ is 3 according to Definition 7.

**Property 1.** Given an itemset $P$ and two subsets of $DB$: $DB_1$ and $DB_2$. If $(DB_1 \cap DB_2 = \emptyset) \wedge (DB_1 \cup DB_2 = DB)$, $SC(P) = SC(P, DB_1) + SC(P, DB_2)$.

**Proof.** Let $DB_P$ be the set of all transactions containing $P$ in $DB$. Because of $DB_1 \cup DB_2 = DB$, we have $DB_P = DB_P \cap DB = DB_P \cap (DB_1 \cup DB_2) = (DB_P \cap DB_1) \cup (DB_P \cap DB_2)$. According to Definition 2 and Definition 7, $SC(P)$ is equal to $|DB_P|$. According to Definition 6 and Definition 7, $SC(P, DB_1)$ and $SC(P, DB_2)$ are equal to $|DB_P \cap DB_1|$ and $|DB_P \cap DB_2|$ respectively. Because of $DB_1 \cap DB_2 = \emptyset$, we have $(DB_P \cap DB_1) \cap (DB_P \cap DB_2) = \emptyset$. That is, $|(DB_P \cap DB_1) \cup (DB_P \cap DB_2)| = |DB_P \cap DB_1| + |DB_P \cap DB_2|$. Therefore, $SC(P) = |DB_P| = |(DB_P \cap DB_1) \cup (DB_P \cap DB_2)| = |DB_P \cap DB_1| + |DB_P \cap DB_2| = SC(P, DB_1) + SC(P, DB_2)$.

The framework of the extensional approaches with TA pruning consists of three parts: (1) maintaining a list $L_k$ to store the top-$k$ co-occurrence items that have been found so far, (2) computing the lower bound of support counts of items in $L_k$ and the upper bound of support counts of all remaining candidate items, and (3) deciding whether the mining process should be terminated according to the lower bound and the upper bound. The only difference between the extensional approaches lies in how to estimate the upper bound.

In the following part, we first introduce how to extend NT with TA pruning and then present the extensional version of NTI.

Given itemset $P$, let $DB_{unvisited}$ be the set of all transactions that are not visited at present and $I_c$ be the set of all candidate items that do not belong to $L_k$ currently. We denote the maximal value of $\{SC(P \cup \{i\}, DB - DB_{unvisited}) \mid i \in I_c\}$ as $MAXV_{lsc}$ and denote the minimal value of $\{SC(P \cup \{i\}, DB - DB_{unvisited}) \mid i \in L_k\}$ as $MINV_{lsk}$. NT-TA, the extensional version of NT, sets the lower bound to $MINV_{lsk}$ and the upper bound to $MAXV_{lsc} + |DB_{unvisited}|$. As for $MINV_{lsk}$ and $MAXV_{lsc}$, we have following properties.

**Property 2.** For any $i \in L_k$, $SC(P \cup \{i\}) \geq MINV_{lsk}$.

**Proof.** Because $MINV_{lsk}$ is the minimal value of $\{SC(P \cup \{i\}, DB - DB_{unvisited}) \mid i \in L_k\}$, we have $SC(P \cup \{i\}, DB - DB_{unvisited}) \geq MINV_{lsk}$ for any $i \in L_k$. In addition, $SC(P \cup \{i\}) \geq SC(P \cup \{i\}, DB - DB_{unvisited})$ holds according to Definition 2 and Definition 6. Therefore, $SC(P \cup \{i\}) \geq SC(P \cup \{i\}, DB - DB_{unvisited}) \geq MINV_{lsk}$.

**Property 3.** For any $i \in I_c$, $SC(P \cup \{i\}) \leq MAXV_{lsc} + |DB_{unvisited}|$.

**Proof.** According to Property 1, $SC(P \cup \{i\}) = SC(P \cup \{i\}, DB - DB_{unvisited}) + SC(P \cup \{i\}, DB_{unvisited})$. Because of $SC(P \cup \{i\}, DB - DB_{unvisited}) \leq MAXV_{lsc}$ and $SC(P \cup \{i\}, DB_{unvisited}) \leq |DB_{unvisited}|$, we have $SC(P \cup \{i\}) \leq MAXV_{lsc} + |DB_{unvisited}|$.

According to Property 2 and Property 3, we have the following Lemma.

**Lemma 1.** If $MINV_{lsk} > MAXV_{lsc} + |DB_{unvisited}|$, any item that is not in $L_k$ can't be a top-k co-occurrence item.

Lemma 1 indicates that current $L_k$ is the final result If $MINV_{lsk} > MAXV_{lsc} + |DB_{unvisited}|$. NT-TA adopts Lemma 1 to early terminate the mining process without scanning all transaction. Algorithm 3 shows the details of NT-TA.

---

**Algorithm 3: NT-TA Algorithm**
**Input:** Transaction database $DB$, itemset $P$, and threshold $k$.
**Output:** $L_k$, the set of all top-$k$ co-occurrence items of $P$.
1: $I_c \leftarrow NULL$; $L_k \leftarrow NULL$;
2: $MINV_{lsk} \leftarrow 0$; $MAXV_{lsc} \leftarrow 0$; $C_u \leftarrow |DB|$;
3: **foreach** $T \in DB$ **do**
4:   **if** $P \subseteq T$ **then**
5:     **foreach** $i \in T - P$ **do**
6:       **if** $i$ is first visited **then**
7:         $CO(i) \leftarrow 1$;
8:       **else**
9:         $CO(i) \leftarrow CO(i) + 1$;
10:       **if** $CO(i) \geq MINV_{lsk}$, **then**
11:         $L_k \leftarrow L_k \cup \{i\}$;
12:         $MINV_{lsk} \leftarrow$ the $k$th biggest value of $\{CO(i) \mid i \in L_k\}$
13:         **foreach** $i \in L_k$ **do**
14:           **if** $CO(i) < MINV_{lsk}$ **then**
15:             $L_k \leftarrow L_k - \{i\}$;
16:       **else**
17:         $I_c \leftarrow I_c \cup \{i\}$;
18:         $MAXV_{lsc} \leftarrow$ the biggest value of $\{CO(i) \mid i \in I_c\}$;
19:   $C_u \leftarrow C_u - 1$;
20:   **if** $MINV_{lsk} > MAXV_{lsc} + C_u$ **then**
21:     **return** $L_k$;
22: **return** $L_k$;

---

In Algorithm 3, initializing variables is implemented by line 1 and line 2, where $C_u$ is used to store the value of $|DB_{unvisited}|$ and $I_c$ maintains the set of candidate items that have an opportunity to become top-$k$ co-occurrence items. In the beginning, no transaction has been visited. Therefore, $C_u$ is set to $|DB|$. From line 3 to line 22, each transaction is examined. If transaction $T$ contains itemset $P$, the co-occurrence count of each item belonging to $T - P$ is increased by 1 (line 5 to line 9). If the co-occurrence count of an item is not less than current $MINV_{lsk}$ (the lower bound), the item is inserted into $L_k$ (line 11), $MINV_{lsk}$ is recomputed (line 12), and $L_k$ is modified to ensure that it only contains top-$k$ co-occurrence items found so far (line 13 to line 15). If the co-occurrence count of an item is less than current $MINV_{lsk}$, it is inserted into $I_c$ for future opportunity (line 17) and $MAXV_{lsc}$ (the upper bound) is recomputed accordingly (line 18). Because the examination of the transaction is finished, $C_u$ is decreased by 1 (line 19). Finally, NT-TA checks whether $MINV_{lsk}$ is greater than $MAXV_{lsc} + C_u$. If $MINV_{lsk}$ is greater than $MAXV_{lsc} + C_u$, the current $L_k$ is sure to contain all final results and NT-TA terminates the mining process (line 20 to line 21).

NTI-TA, the extensional version of NTI with TA pruning, adopts a mining process that is almost the same as that of NT-TA. The only difference is that $MAXV_{lsc}$ is defined as the maximal value of $\{SC(P \cup \{i\}, DB_P - DB_{P,unvisited}) \mid i \in L_k\}$, where $DB_P$ is the set of all transactions containing itemset $P$ and

$DB_{P,unvisited}$ is the set of all transactions, which are not visited at present, in $DB_P$. Replacing $DB_{unvisited}$ with $DB_{P,unvisited}$, the Lemma 1 still holds. That is, If $MINV_{lsk} > MAXV_{lsc} + |DB_{P,unvisited}|$, any item that is not in $L_k$ can't be a *top-k co-occurrence item*. In fact, NTI-TA first generates $DB_P$ by the same way as NTI. Then, it employs the same process as NT-TA to find the set of all top-*k* co-occurrence items just by replacing *DB* with $DB_P$. Algorithm 4 shows the details.

---

**Algorithm 4: NTI-TA Algorithm**
**Input:** Transaction database *DB*, itemset *P*, and threshold *k*.
**Output:** $L_k$, the set of all top-*k* co-occurrence items of *P*.
1: generating $DB_P$ by running line 1 to line 14 of Algorithm 2;
2: generating and return $L_k$ by running Algorithm 3 with replacing *DB* by $DB_P$.

---

## V. APPROACHES BASED ON PI-TREE

In this section, we first introduce the structure of Pi-Tree. Then, based on Pi-Tree, we propose two algorithms to fast discovery all top-*k* co-occurrence itemsets.

### A. Pi-Tree: Definition and Construction

To facilitate the mining efficiency, we use a compact structure named *Pi-Tree* to maintain the information of items in a database.

**Definition 8.** A Pi-Tree is a prefix tree structure defined below.
(1) It is made up of one root labeled as "*Root*", a set of item prefix subtrees as the children of the root, and a header table.
(2) Each node consists of four fields: *label*, *count*, *children-link*, and *parent-link*. Field *label* registers which item this node represents. Field *count* registers the number of transactions presented by the portion of the path reaching this node. Field *children-link* registers all children of the node. Field *parent-link* registers the parent of the node.
(3) Each entry in the header table consists of two fields: *label* and *node-link*. Field *label* registers items. Field *node-link* registers a list of nodes with the corresponding *label*. In the header table, entries are sorted in the descending order of support count. Header table is employed to facilitate the traversal of Pi-Tree.

In terms of the above definition, a Pi-Tree looks like a FP-tree [11]. However, there are two important differences between them. First, Pi-Tree doesn't require items registered in nodes to be frequent while FP-tree only focuses on frequent items. Second, Nodes in a Pi-Tree have field parent-link while Nodes in a FP-tree don't have such field. Field parent-link has been used to incrementally mine frequent itemsets and proved to be efficient [19]. FP-tree can be regarded as a special kind of Pi-Tree.

Given a database, a Pi-Tree can be constructed with only two scans of the database. In the first scan, the support count of each item is computed. In the second scan, a tree with root *Root* is first created. Subsequently, transactions are reorganized and inserted into the Pi-Tree one by one. Whenever a transaction is retrieved, items in the transaction are sorted in descending order of support count and then inserted into the Pi-Tree sequentially. Algorithm 5 shows the details.

---

**Algorithm 5: Pi-Tree Construction**
**Input:** A transaction database *DB*
**Output:** *PTr*: a Pi-Tree
1: Scan *DB* to find all items and their support counts;
2: initialize a Pi-Tree, *PTr*, with a header table;
3: **foreach** transaction *T* in *DB* **do**
4:     sorting *T* in descending order of support count;
5:     $i \leftarrow T.first\text{-}item$; //*T.first-item* is the first item of sorted *T*
6:     $T \leftarrow T - \{i\}$;
7:     **Call** *Insert_Tree*(*i*, *T*, *PTr*);
8: **return** *PTr*;

---

The function *Insert_Tree* (*i*, *T*, *PTr*) is performed as follows. If *PTr* has a child *N* such that *N.label = i*, then increase *N*'s *count* by 1; else create a new node *N*, with its *count* initialized to 1, and add it to *PTr*'s children-link. Meanwhile, the new node *N* is added to the *node-link* of item *i*. If *T* is nonempty, call *Insert_Tree*(*T.first-item*, *T* − {*T.first-item*}, *N*) recursively. Figure 1 shows the Pi-Tree for Table 1 after running Algorithm 5. Notice that each node is assigned a name like $M_j$ for identification purpose.

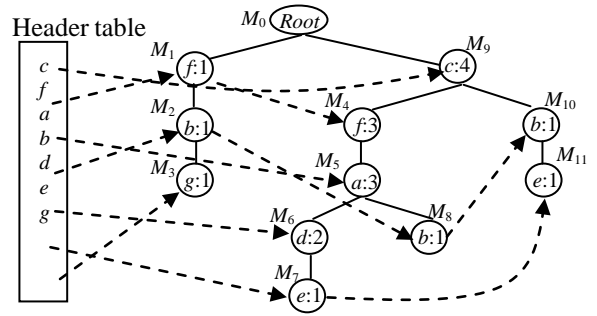

Figure 1. The Pi-Tree for Table 1

### B. Pi-Tree: Definition and Construction

Definition 9. Given a database, we define $>_{dsc}$, an order among all items, as descending order of support count. For two items i1 and i2, i1 $>_{dsc}$ i2 if and only if i1 is ahead of i2 according to $>_{dsc}$.

For example, the order $>_{dsc}$ for Table 1 is {*c, f, a, b, d, e, g*}. Because *a* is ahead of *e* in the order, we have $a >_{dsc} e$. Note that, if the support counts of some items are equal, the orders among these items can be assigned arbitrarily.

For the sake of our discussion, the following conversion holds in the remainder of this paper.

**Conversion 1**. *An itemset is sorted by $>_{dsc}$ order when*

*mentioned.*

**Property 4.** Let $i_1$ and $i_2$ be two items. $Nd_1$ and $Nd_2$ are two nodes registering $i_1$ and $i_2$ respectively. If $i_1 >_{dsc} i_2$, $Nd_2$ can not be an ancestor of $Nd_1$.

**Proof.** Assume $Nd_2$ is an ancestor of $Nd_1$. According to the construction of Pi-Tree, there must have a transaction containing $i_1$ and $i_2$. When this transaction is inserted into the Pi-Tree, $i_1$ and $i_2$ are registered by $Nd_1$ and $Nd_2$. Because $Nd_2$ is an ancestor of $Nd_1$, $i_2$ must be processed ahead of $i_1$. However, according to the procedure of inserting transactions into a Pi-Tree, the transaction is first sorted in order $>_{dsc}$ and then the items in the sorted transaction are registered in nodes sequentially. This means that $i_1$ must be processed ahead of $i_2$. Therefore, the assumption is wrong.

**Property 5.** Let $Nd_1$ and $Nd_2$ are two nodes, which register $i_1$ and $i_2$ respectively. If $Nd_1$ and $Nd_2$ are in the same path and $i_1 >_{dsc} i_2$, $Nd_1$ must be an ancestor of $Nd_2$.

**Proof.** If $Nd_1$ and $Nd_2$ are in the same path, either $Nd_1$ is an ancestor of $Nd_2$ or $Nd_2$ is an ancestor of $Nd_1$. According to Property 4, $Nd_2$ can not be an ancestor of $Nd_1$ because of $i_1 >_{dsc} i_2$. Therefore, $Nd_1$ must be an ancestor of $Nd_2$.

**Definition 10.** A path is called a *full path* if it is a path from a leaf to the root.

**Definition 11.** Given an itemset $P$ and a path $Pa$, $Pa$ is said to *register P* if for any $i$ in $P$, $Pa$ has a node registering $i$.

**Definition 12.** Given itemset $P$ and item $i$ ($\in P_{coi}$), $SFP(P, i)$ is defined as the set of all full paths registering $P \cup \{i\}$.

**Definition 13.** The register node set of $i$, denoted as $NR(i)$, is the set of all node registering $i$.

**Lemma 2.** Let $P$ be an itemset and $i$ be an item belonging to $P_{coi}$. The last item of $P$ is denoted as $i_{LoP}$. The following conclusions hold.

(1) If $i_{LoP} >_{dsc} i$, $CO(P, i)$ is the sum of *counts* of all nodes that register $i$ and are in a full path belonging to $SFP(P, i)$, i.e.,

$$CO(P,i) = \sum_{(N \in NR(i) \wedge (\exists Pa \in SFP(P,i), N \in Pa.Nset)} N.count \; ,$$

where $Pa.Nset$ represents the set of all nodes in path $Pa$.

(2) If $i >_{dsc} i_{LoP}$, $CO(P, i)$ is the sum of *counts* of all nodes that register $i_{LoP}$ and are in a full path belonging to $SFP(P, i)$, i.e.,

$$CO(P,i) = \sum_{(N \in NR(i_{LoP})) \wedge (\exists Pa \in SFP(P,i), N \in Pa.Nset)} N.count \; .$$

Proof. We first prove conclusion (1).

For simplicity, we denote the path from a node $N$ to the root as $Path(N)$, and $\{N \mid (N \in NR(i)) \wedge (\exists Pa \in SFP(P, i), N \in Pa.Nset)\}$ as $SCN_{P,i}$. Without loss of generality, we assume $SCN_{P,i} = \{N_1, N_2, ..., N_s\}$. Let $DB_j$ be the set of all transactions that register $i$ to $N_j$. According to the construction algorithm of Pi-Tree, $|DB_j|$ is equal to $N_j.count$. Obviously, $DB_j \cap DB_l$ ($j \neq l$) is null because each transaction is just inserted into the Pi-Tree once.

Notice that $N_j$ must satisfy constraint "$\exists Pa \in SFP(P, i), N_j \in Pa.Nset$". That is, $N_j$ must be a node in a full path registering $P \cup \{i\}$. We denote the full path as $Pa_j$. Let $x$ be an item in $P$.

According to Conversion 1, $x >_{dsc} i_{LoP}$ holds if $x$ is not $i_{LoP}$. As a precondition of conclusion (1), $i_{LoP} >_{dsc} i$ holds. Therefore, $x >_{dsc} i$ holds. This means that the node registering $x$, denoted as $N_x$, in $Pa_j$ is an ancestor of $N_j$ according to Property 5. That is, $N_x$ is in the path from the root to $N_j$. According to the procedure of inserting transactions into the Pi-Tree, any transaction, denoted as $T$, registering $i$ to $N_j$ must contains $x$. That is, $T$ should be a superset of $P$. Therefore, $DB_j$ can be rewritten as $\{T \mid (T \in DB) \wedge (P \cup \{i\} \subseteq T) \wedge (T \text{ registers } i \text{ to } N_j)\}$.

Let $DB_P = \{T \mid (T \in DB) \wedge (P \cup \{i\} \subseteq T)\}$ be the set of all transactions containing itemset $P \cup \{i\}$. According to Definition 4 and Definition 7, $CO(P, i)$ is equal to $|DB_P|$.

Now, we prove that $\cup_{1 \leq j \leq s} DB_j$ is equal to $DB_P$. According to the rewritten definition of $DB_j$, $DB_j$ should be a subset of $DB_P$. Therefore, we have $\cup_{1 \leq j \leq s} DB_j \subseteq DB_P$. On the other hand, for any transaction, denoted as $T_P$, in $DB_P$, $T_P$ must register $i$ to a node, denoted as $N_P$, in the Pi-Tree. Let $Pa_f$ be the full path containing $N_P$. Because of $P \cup \{i\} \subseteq T$, $Pa_f$ must contains these nodes, which register each item of $P \cup \{i\}$ respectively. That is, $Pa_f$ is a full path registering $P \cup \{i\}$. Therefore, $Pa_f \in SFP(P, i)$. According to the definition of $SCN_{P,i}$, $N_P$ must be an element of $SCN_{P,i}$. thus, $N_P$ should be one of $N_1, N_2, ..., $ and $N_s$. Let $N_P$ is $N_l$. According to the rewritten definition of $DB_l$, $T_P$ should be an element of $DB_l$. This means $DB_P \subseteq \cup_{1 \leq j \leq s} DB_j$. Therefore, $\cup_{1 \leq j \leq s} DB_j$ is equal to $DB_P$. Because of $DB_u \cap DB_v$ ($u \neq v$), $|DB_P| = |\cup_{1 \leq j \leq s} DB_j| = \sum_{1 \leq j \leq s} |DB_j|$ holds. Based on the above analysis, conclusion (1) holds.

In the same way, we can prove conclusion (2).

Lemma 2 indicates that $SFP(P, i)$ includes whole information about co-occurrence items of $i$. Therefore, the problem of mining top-$k$ occurrence items is transformed into the problem of finding $SFP(P, i)$. In the next subsection, we will explore how to find $SFP(P, i)$ efficiently.

*C. PT Algorithm*

Before introducing the PT algorithm, let's first examine an example. Assume we want to find the top-2 co-occurrence items of {ca} from Table 1. Lemma 2 indicates that co-occurrence counts of {ca} and its co-occurrence items can be directly computed from the full paths registering item c and item a. In Figure 1, only two full paths, M0M9M4M5M6M7 and M0M9M4M5M8, register a and c. Because of f >dsc a, CO({ca}, f) is equal to the count of node M5 according to Lemma 2. That is, CO({ca}, f) = 3. Because of a >dsc d, CO({ca}, d) is equal to the count of node M6. That is, CO({ca}, d) = 2. Similarly, CO({ca}, b) = 1 and CO({ca}, e) = 1. Therefore, the top-2 co-occurrence itemsets of {ca} are f and d. By carefully examining the example, we find all co-occurrence items of {ca} are contained in only two parts. The first part is the path from node M5 to the root. The second part is the subtree rooted at M5. Of course, M5 must satisfy the condition that the path from it to the root must register {ca}. Therefore, our strategy is described as following. First, we locate each node registering a. Second, we find these paths from nodes registering a to the root, where some node register c. Meanwhile, we obtain some co-occurrence items of {ca}. Then, for each above path, we scan the subtree rooted at the node registering a to get all other co-occurrence items and their co-occurrence counts. Finally, we find the top-2 co-occurrence items by sorting these items in descending order of co-occurrence count.

Given an itemset $P$, the core problem is how to fast find these nodes registering $i_{LoP}$ with the paths from them to the root registering $P$.

**Property 6.** Let $i_1$, $i_2$, and $i_3$ be three items with $i_1 >_{dsc} i_2 >_{dsc} i_3$. Assume node $N_1$ and node $N_3$ register $i_1$ and $i_3$ respectively and there exist a path named Path$_{31}$, from $N_3$ to $N_1$. Let Path$_{30}$ be the path from $N_3$ to the root. If no node in Path$_{31}$ registers $i_2$, no node in Path$_{30}$ can register $i_2$.

**Proof.** According to the construction of Pi-Tree, $N_1$ must be in Path$_{30}$. In fact, Path$_{30}$ can be divided into two parts: Path$_{31}$ and Path$_{30}$ − Path$_{31}$. According to the precondition, no node in Path$_{31}$ registers $i_2$. Let $N$ be a node in Path$_{30}$ − Path$_{31}$ and registers $i_k$. According to the construction of Pi-Tree, we have $i_k >_{dsc} i_1$. That is, no node in Path$_{30}$ − Path$_{31}$ registers $i_2$ because of $i_k >_{dsc} i_1 >_{dsc} i_2$. Therefore, no node in Path$_{31}$ or Path$_{30}$ − Path$_{31}$ can registers $i_2$.

---

**Algorithm 6: PT Algorithm**
**Input:** an itemset $P = i_1 i_2 \ldots i_s$ and a Pi-Tree $PTr$
**Output:** $L_k$, the set of all top-$k$ co-occurrence items of $P$.
1: Find $CNS$, the set of all nodes which register $i_s$, by accessing the entry of $i_s$ in the header table of $PTr$.
2:   $NS \leftarrow \varnothing$;
3: **foreach** node $N$ in $CNS$ **do**
4:     $x \leftarrow s - 1$;
5:     $flag \leftarrow$ false;
6:     $Current\_node \leftarrow N.father$;
7:     **do**
8:       **if** $i_x = Current\_node.label$ **then**
9:         $x \leftarrow x - 1$;
10:         $Current\_node \leftarrow Current\_node.father$;
11:       **else**
12:         **if** $i_x >_{dsc} Current\_node.label$ **then**
13:           $Current\_node \leftarrow Current\_node.father$;
14:         **else**
15:           $flag \leftarrow$ true;
16:     **until** $(x = 0) \vee (flag =$ true$)$
17:     **if** $flag =$ false **then**
18:       $NS \leftarrow NS \cup \{N\}$;
19: **foreach** Node $N$ in $NS$ **do**
20:     scan the path from $N$ to $root$. For node $Nd$ in the path, if the label of $Nd$ is not in $P$, increase $CO(P, Nd.label)$ by $N.count$.
21:     travel the subtree rooted at $N$, for each no-root node $Nd$, increase $CO(P, Nd.label)$ by $Nd.count$.
22: sort all $CO(P, i)$ by descending order of value, and assign the set of top-$k$ co-occurrence items to $L_k$.
23: **return** $L_k$.

---

Combining our strategy with Property 6, we propose PT as described by Algorithm 6. Variable $flag$ is used to indicate whether the precondition of Property 6 is satisfied. $flag$ is set to true when the precondition of Property 6 is satisfied. According to Property 6, when $flag$ becomes true, the current path is undesirable and need not be further processed. If $flag$ is unchanged after all nodes of a path are visited, the path must register $P$ and the original node initiating the path is desirable.

In Algorithm 6, Line 1 uses $CNS$ to store all nodes which register $i_s$, the last item of itemset $P$. Line 2 initializes $NS$, the set of nodes which register $i_s$ and are desirable for finding all top-$k$ co-occurrence items of itemset $P$, to be null. Line 3 to 16 find such desirable nodes from $CNS$ by checking whether each item in $P$ except $i_s$ is registered by a node in the path from current node N, registering $i_s$, to the root. Line 15 employs Property 6 to void unnecessarily comparison since $Current\_node.label >_{dsc} i_x$, which means no node of the path from current node $N$ to the root can register $i_x$. That is, node $N$ is undesirable and no further comparison should be performed. For each item $i$ ($\notin P$) which co-occurs with itemset $P$, Line 19 to 21 compute the *co-occurrence count* of $P$ and $i$, namely $CO(P, i)$. Line 22 finally finds all top-$k$ co-occurrence items of itemset $P$ by sorting all $CO(P, i)$ by descending order of value.

### D. PT-TA Algorithm

**Definition 14**. Let i be an item and N be a node. We define fre(i, N) be the sum of counts of all nodes registering i in the subtree rooted at N.

**Lemma 3.** Let $P$ be itemset and $i$ be an item belonging to $P_{coi}$. The last item of $P$ is denoted as $i_{LoP}$. Let $SCN_{P,i} = \{N_1, N_2, \ldots, N_t\}$ be the set of all nodes registering $i_{LoP}$ with the constraint that the paths from these nodes to the root register $P$. If $i_{LoP} >_{dsc} i$, we have

$$CO(P,i) = \sum_{1 \le j \le t} fre(i, N_j).$$

Lemma 3 can be easily inferred from Lemma 2. Limited by space, we omit the proof. According to the construction of Pi-Tree, fre(i, N) is not bigger than N.count. Therefore, we have the following lemma inferred from Lemma 3 directly.

**Lemma 4.** For any x ($1 \le x \le t$), we have

$$CO(P,i) \le \sum_{1 \le j \le x} fre(i, N_j) + \sum_{x+1 \le j \le t} N_j.count.$$

Replacing MAXVlsc + |DBunvisited| with max $\{\Sigma 1 \le j \le x$ fre(i, Nj) | i∈ Ic$\} + \Sigma$ x+1 $\le j \le$ t Nj.count, where Ic be the set of candidate items, Lemma 1 still holds. By setting the upper bound to max $\{\Sigma 1 \le j \le$ x fre(i, Nj) | i∈ Ic$\} + \Sigma$ x+1 $\le j \le$ t Nj.count, we adopt a mining process similar to NT-TA to extend PT Algorithm by using TA pruning strategy. Algorithm 7 shows the details.

In Algorithm 7, Line 1 finds the desirable nodes by the same method used in Algorithm 6. Line 2 initializes $L_k$, which stores top-$k$ co-occurrence items, and $C_k$, which stores candidate items, to be null. Line 3 to 5 find all top-$k$ co-occurrence items from all items that co-occur with itemset $P$ and are ahead of $i_s$, the last item of itemset $P$, according to $>_{dsc}$. Line 7 to 16 find the finally top-$k$ co-occurrence items by checking all items which are behind $i_s$ according to $>_{dsc}$ and co-occur with itemset $P$. Line 15 employs Lemma 1 to avoid unnecessary computation.

---

**Algorithm 7: PT-TA Algorithm**
**Input:** an itemset $P = i_1\ i_2…i_s$ and a Pi-Tree $PTr$
**Output:** $L_k$, the set of all top-$k$ co-occurrence items of $P$.
1: generate $NS = \{N_1, N_2, …, N_t\}$ by line 1 to 17 of Algorithm 6;
2: $L_k \leftarrow \varnothing$; $I_c \leftarrow \varnothing$;
3: **foreach** Node $N$ in $NS$ **do**
4:   scan the path from $N$ to *root*. For node $Nd$ in the path, if the label of $Nd$ is not in $P$, increase $CO(P, Nd.label)$ by $N.count$.
5: sort all above $CO(P, i)$ by descending order of value, assign all items with top-$k$ values to $L_k$, and add residual items to $I_c$.
6: $lowerbound \leftarrow \min\{CO(P, i) \mid i \in L_k\}$;
7: **for** $j = 1$ to $t$ **do**
8:   **foreach** node $Nd$ in the subtree rooted at $N_j$ **do**
9:     increase $CO(P, Nd.label)$ by $Nd.count$;
10:     **if** $CO(P, Nd.label) \geq lowerbound$ **then**
11:       insert $Nd.label$ into $L_k$;
12:       remove items, whose $CO$ value are not the biggest top-$k$, from $L_k$ to $I_c$.
13:       recompute $lowerbound$ according to the modified $L_k$;
14:   $upbound \leftarrow \max\{CO(P, i) \mid i \in I_c\} + \sum_{j+1 \leq u \leq t} N_u.count$;
15:   **if** $upbound < lowerbound$ **then**
16:     **exit** from **for loop** and **return** $L_k$;
17: **return** $L_k$.

---

## VI. EXPERIMENTS AND RESULTS

In this section, we evaluate the performance of all six algorithms. The experiments were performed on a PC server with 16G memory and 2GHZ Intel processor. All codes were implemented by C# and ran on X64 windows server 2003 system.

### A. Databases

Two real-world databases and two synthetic databases are used in our experiments. Table 2 shows the characteristics of these databases. In the last column of Table 2, we use transaction count divided by item count to provide a simple way to measure density of databases. Larger value means a dense database. According to this measurement, Syn_data1 is the densest one.

Connect and Accidents are real-world databases. They are often used in previous study of frequent itemset mining. To test our algorithms' performance on large-scale databases, we generated large synthetic databases by IBM Quest Synthetic Data Generator. The synthetic database named Syn_data1 and Syn_data2 is used in our experiments. To generate Syn_data 1, the average transaction size, average maximal potentially frequent itemset size, and correlation between patterns are set to 40, 30, and 0.25 respectively, while the number of transactions, number of different items, and number of patterns are set to 1000K, 1K, and 10 respectively. The parameters for generating Syn_data2 are the same as these for generating Syn_data1 except that the number of patterns is set to 100, which makes Syn_data2 sparser than Syn_data1.

In the following subsections, we will show and analyze experiment results on different metrics including preprocessing time, memory used, running time, and scalability.

**Table 2. Characteristics of databases used in our experiments**

| Database | Transaction Count | Item Count | Avg. Length | Density |
|---|---|---|---|---|
| Connect | 67,557 | 129 | 43 | 524 |
| Accidents | 340,183 | 468 | 34 | 727 |
| Syn_data1 | 1,000,000 | 198 | 42 | 5,051 |
| Syn_data2 | 1,000,000 | 687 | 41 | 1,456 |

### B. Preprocessing Time and Memory Usage

After databases are loaded into memory, NTI needs to build Tid-sets and PT needs to build a Pi-tree. In this section we will show performance of their preprocessing phases. The preprocessing phase begins when a database is loaded into memory and ends when the algorithm is ready to run to discover the results of given query itemsets. Note that, since the preprocessing phase of these algorithms and their extensional versions with TA pruning are the same, we ignore the discussion on these extensional versions.

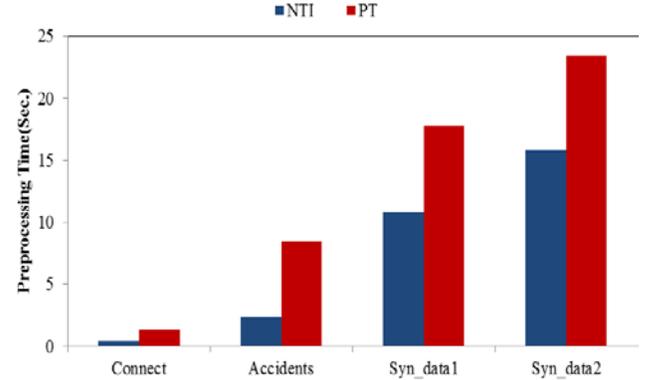

**Figure 2. Preprocessing time**

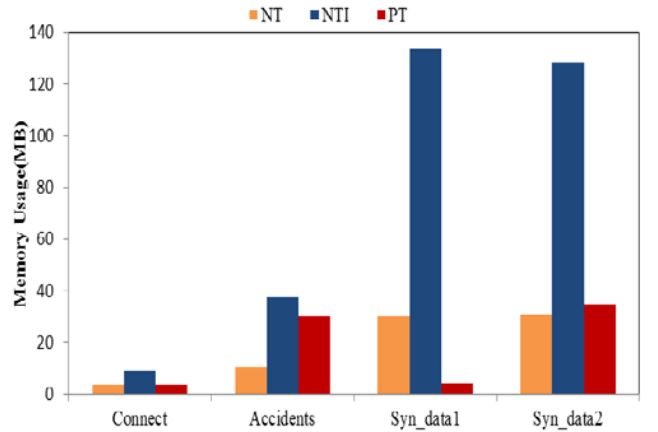

**Figure 3. Preprocessing memory usage**

First, we evaluate preprocessing time of each algorithm. The results are shown in Figure 2. Since NT doesn't have index phase, we ignore the discussion on it. From Figure 2, we observe that the preprocessing time of PT is longer on all databases than that of NTI. This is because NTI only transposes the data as Tid-sets, while PT has to reorganize transactions to build a Pi-tree, which is more time-consuming. The detailed experimental data shows that the preprocessing time of PT is about 3.5 times longer than NTI on real-world databases and about 1.5 times longer than NTI on synthetic databases. From this observation, we find that the more dense databases are, the smaller the difference of preprocessing time between NT and PT is.

Second, we compare each algorithm's memory usage in the preprocessing phase. The result is shown in Figure 3. From Figure 3, we observe that the memory usage of PT is very large on Accidents. The reason is that Accidents have many items which lead to generating large numbers of nodes. There are about 5 million nodes in the Pi-tree of Accidents while about half a million nodes in other three databases. We also observe that Pi-tree needs much less memory than Tid-set on dense databases. The reason is that Pi-tree has the functionality of compressing data and the compress rate is high when data is dense, while Tid-set only transposes data so its memory usage is strongly related to databases' size. It should be noticed that at the mining phase, NT and NTI still needs to scan the original database, while PT doesn't need to access the original database any more after indexing it. Therefore, the memory usage of PT is much better than that of NT and NTI when database is dense and huge, such as Syn_data1.

### C. Running Time

In this section, we evaluate the running time of each algorithm. We generate query itemsets with different length from 3 to 7 for each database. For each query itemset with length of x and each database, we generate 100 query itemsets and all the algorithms find the top-k co-occurrence items for these itemsets. To avoid query itemsets with no results, we generate queries in the following process: (1) Randomly select a transaction in the database; (2) Randomly pick x items in this transaction to form a query itemset.

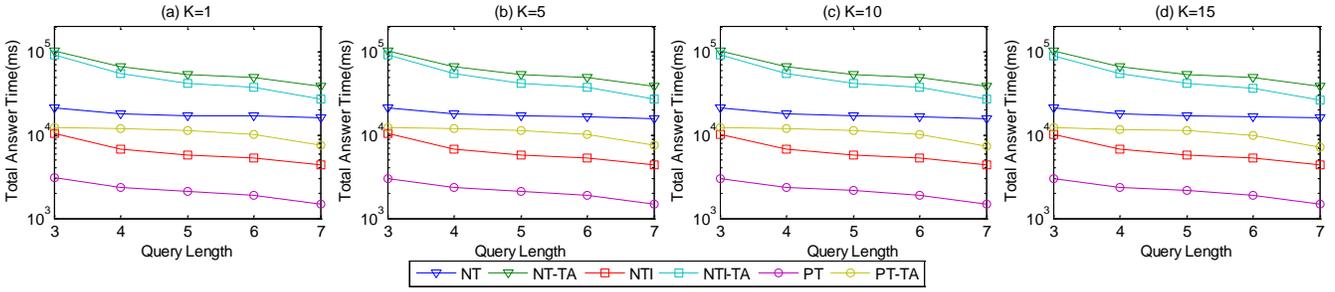

**Figure 4. Running time on Connect**

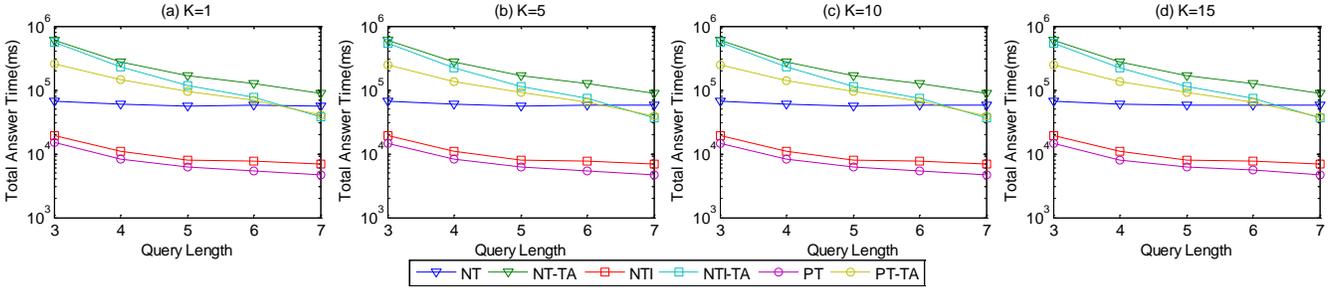

**Figure 5. Running time on Accidents**

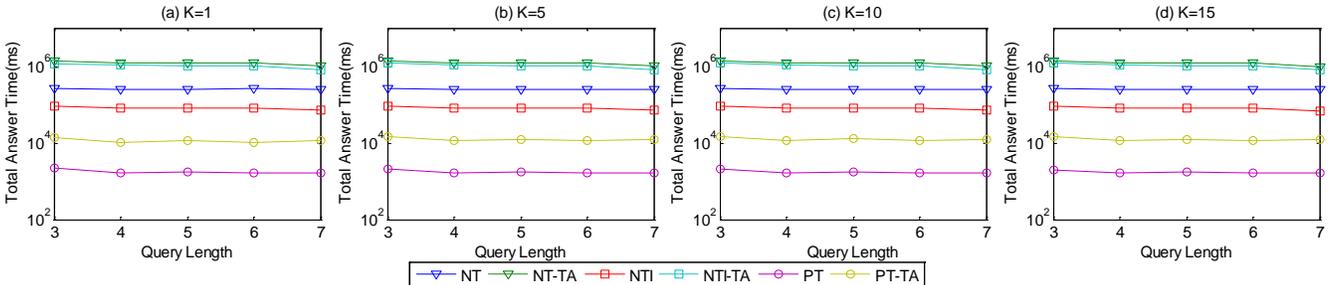

**Figure 6. Running time on Syn_data1**

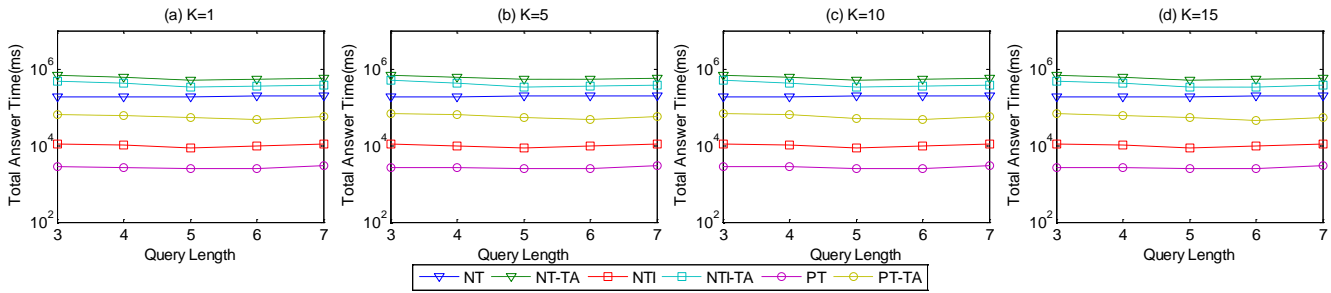

Figure 7. Running time on Syn_data2

Sometimes a query itemset will have more than k results, because there is more than one item having the same support as the kth item. In this case, all the algorithms will return answers containing all the items having the same support as the kth item even though the result is longer than k.

First, let's see the results of Connect in Figure 4. PT outperforms NTI by only using one third time to mining results for all query itemsets, and NTI beats NT. Tid-set, the inverted index adopted by NTI, accelerates the process of seeking valid transactions, therefore NTI is faster than NT. Similar transactions will be under the same branches in a Pi-tree. Therefore, for the same query itemsets, the number of valid nodes in a Pi-tree is usually much less than the number of valid transactions found by NTI. In our experiment, PT finds about 9,511 valid nodes per query itemset on average while NTI finds about 63,327 valid transactions per query itemset on average. As a consequence, PT is faster than NTI. We also observe that all extensional algorithms with TA pruning are worse than their original versions. The reason may be caused by the following two factors. First, the maintenance of the lower bound and upper bound is time-consuming. Second, the lower bound and upper bound used by these algorithms are too loose to help them terminate early enough. Another fact that we observe is that as the query itemset gets longer, it takes less time for these algorithms to find the results except NT. The reason can be interpreted as following. The number of valid nodes or transactions is smaller when the query itemset becomes longer. For example, average 6,678 valid nodes are found for query itemsets with length of 7 while average 10,512 valid nodes are found for query itemsets with length of 3. However, NT has nothing to do with the length of query itemsets in that it always traverses all the transactions of a database.

The results on Accidents are shown in Figure 5. PT is still faster than NTI, but their performances are close. The time spent by PT is about two-thirds of that spent by NTI. PT finds about 33,468 valid nodes and NTI finds 82,305 valid transactions on average. Compare with corresponding numbers on Connect, these numbers are closer. The reason is that Accidents has more items which lead to an exponential increase of nodes. This also leads to more distinct difference between original algorithms and corresponding extensional versions as finding interim top-$k$ co-occurrence items needs more time when the number of items is larger.

Figure 6 shows the results on Syn_data1, which is the densest one among all databases. On this database, PT is most efficient and is about 40 times faster than NTI. PT finds about 7,527 valid nodes per query itemset on average while NTI finds about 642,772 valid transactions per query itemset on average. There are only 585,720 nodes in the Pi-tree of Syn_data1. The Pi-tree is almost the same size as the Pi-tree of Connect even though the size of Syn_data1 is 20 times larger than the size of Connect. An interesting observation is that as the query itemset gets longer, the mining time doesn't become shorter markedly as previous. This is caused by the density of databases. PT finds about 8,856 valid nodes for query itemsets with length of 3 and finds about 7,219 valid nodes for query itemsets with length of 7 on average. In a very dense database, a lot of items often occur together in transactions.

At last, the results on Syn_data2 are shown in figure 7. As mentioned in Subsection 5.1, this database has the same number of transactions as Syn_data1 while it is sparser than the latter. We can observe that PT is still the fastest one. The runtime of PT on Syn_data2 is 5 times faster than that of NTI. In our experiment, PT finds about 9,965 valid nodes per query itemset on average while NTI finds about 64,668 valid transactions per query itemset on average. We also observe that all extensional algorithms are worse than their original versions as found in the results on Connect, Accidents, and Syn_data1.

From Figure 4 to Figure 7, we find PT outperforms other algorithms substantially and all extensional algorithms are slower than the corresponding original algorithms. The reason may be that the bounds for terminate condition is too loose. We will seek tighter bounds in the future work. NT is the slowest algorithm and performs very stable among three original algorithms. The reason is that NT always traverses all the transactions and its performance is only related to the size of databases. Comparing the performance of PT on all four databases, we find that PT performs better on dense databases and also performs better on databases containing less items when the density doesn't change.

Although the quantitative analysis of the time complexity is helpful to understand the excellent performance of PT, it is very hard to do so since the pruning strategies used in PT depend on the data distribution of a database. As we know, the data distributions of different databases are various and no general distribution can cover all databases. Therefore, the quantitative analysis of PT based on data distributions is intractable. However, we will try to do it in future work.

## D. Scalability

Finally, we evaluate the scalability of these algorithms in this section. Because NT and all extensional algorithms with TA pruning are too slow when databases are large, we only compare the scalability of NTI and PT. We choose Syn_data1 as the baseline database to conduct scalability test. We generate new synthetic databases with the same parameter setting as Syn_data1 except that the number of transactions is set to 2M, 3M, 4M and 5M respecivley. As we observe from Figure 4 to 7, the performance difference between PT and NTI is similar when the query itemsets' length and threshold k changes, we only provide the experiment results by setting length to 5 and k to 10. Figure 8 shows the scalability results. In Figure 8, the Y-axis represents the ratio between the time spent on these larger databases and the time spent on Syn_data1. The X-axis represents the number of transactions in each database. From Figure 8, we observe that as databases get larger, the running time of NTI grows faster than that of PT. This means that PT has better scalability than NTI. This is due to the compressing functionality of Pi-tree. If the distribution of data remains unchanged, when the database becomes large, the size of Pi-tree changes little.

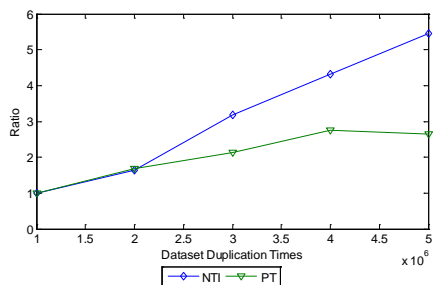

**Figure 8. Scalability on large databases**

## VII. CONCLUSION AND FUTURE WORK

In this paper, we focus on a new mining task named top-$k$ co-occurrence item mining. To address this issue, we have proposed six algorithms for mining top-$k$ co-occurrence items from databases. These algorithms are NT, NTI, PT, NT-TA, NTI-TA, PT-TA. NT-TA, NTI-TA, PT-TA are the extensions of NT, NTI, and PT by using TA strategy to prune the search space. By employing a compressed Pi-Tree to store a database, PT incorporates some novel strategies to find top-$k$ co-occurrence items from the Pi-Tree, thus greatly pruning the search space and achieving high efficiency. In the experiments, different types of synthetic and real databases are used to evaluate the performance of the six proposed algorithms. The experimental results show that PT not only outperforms other algorithms substantially in terms execution time but also has excellent scalability.

As indicated by our experiments, algorithms with TA pruning perform worse than algorithms without TA pruning. This may result from that the bounds for terminate condition is too loose. Therefore, exploring tighter bounds to promote the efficiency of algorithms with TA pruning is our future work. In addition, interactive top-$k$ co-occurrence item mining, the problem of efficiently mining top-$k$ co-occurrence items with changed $k$, is also an interesting topic for future research. Finally, it is also an interesting and challenging work to quantitatively analyze the time complexity of PT.

ACKNOWLEDGEMENT

This work is partially supported by Project 61170091 supported by National Natural Science Foundation of China and Project 2009AA01Z136 supported by the National High Technology Research and Development Program of China (863 Program).